\documentclass{aastex}
\usepackage{emulateapj5}

\newif\ifAMStwofonts
\AMStwofontstrue

\def\kms{km~s$^{-1}$}
\def\apss{Ap\&SS}
\def\degree{$^{\circ}$}

\def\ga{\mathrel{\hbox{\rlap{\hbox{\lower4pt\hbox{$\sim$}}}\hbox{$>$}}}}
\def\la{\mathrel{\hbox{\rlap{\hbox{\lower4pt\hbox{$\sim$}}}\hbox{$<$}}}}

\shorttitle{The disk/halo interface region in the outer Galaxy}

\shortauthors{S.\ Stanimirovi\'{c} et al.}

\begin{document}

\title{First results from the Arecibo Galactic HI Survey: the 
disk/halo interface region in the outer Galaxy}
\submitted{To appear in {\em The Astrophysical Journal}}
\author{Sne\v{z}ana Stanimirovi\'{c}\altaffilmark{1}, Mary
  Putman\altaffilmark{2}, Carl Heiles\altaffilmark{1}, 
  Joshua E. G. Peek\altaffilmark{1}, Paul
  F. Goldsmith\altaffilmark{3,4}, Bon-Chul Koo\altaffilmark{5}, 
  Marko Kr\v{c}o\altaffilmark{3}, 
  Jae-Joon Lee\altaffilmark{5}, Jeff Mock\altaffilmark{6}, 
  Erik Muller\altaffilmark{7}, Jagadheep D. Pandian\altaffilmark{3}, 
  Aaron Parsons\altaffilmark{6}, Yvonne Tang\altaffilmark{3}, 
  Dan Werthimer\altaffilmark{6}}
\altaffiltext{1}{Radio Astronomy Lab, UC Berkeley, 601 Campbell Hall,
Berkeley, CA 94720 sstanimi$@$astro.berkeley.edu}
\altaffiltext{2}{University of Michigan, Department of Astronomy, 500 Church
St., Ann Arbor, MI 48109; mputman$@$umich.edu}
\altaffiltext{3}{Department of Astronomy,
Cornell University, Ithaca, NY 14853}
\altaffiltext{4}{Jet Propulsion Laboratory, 4800 Oak Grove Drive, 
 Pasadena CA 91109} 
\altaffiltext{5}{Department of Physics and Astronomy, Seoul 
National University, Seoul 151-742, Korea}
\altaffiltext{6}{Space Sciences Laboratory, University of California, Berkeley,
  CA 94720}
\altaffiltext{7}{CSIRO Australia Telescope National Facility, PO Box 76, Epping, 
NSW 1710, Australia}

\begin{abstract}
The consortium for Galactic studies with the Arecibo
L-band Feed Array (ALFA) is conducting a neutral hydrogen (HI) 
survey of the whole Arecibo sky (declination range from $-1$\degree~to
38\degree), with high angular (3.5$'$) and velocity 
resolution (0.2 \kms). 
The precursor observations with ALFA of a region in
the Galactic anti-center reveal numerous isolated, 
small (a few pc in size), and cold ($T_{\rm k}<400$ K) HI clouds 
at low negative velocities, distinctly separated from the 
HI disk emission (`low-velocity clouds', LVCs). 
These clouds are most likely located in the transition region between 
the Galactic disk and halo (at scale heights of 60--900 pc), 
yet they have properties of typical cold neutral clouds. 
LVCs are colder and, most likely, smaller and less
massive than Lockman's clouds in the disk/halo interface region of 
the inner Galaxy. 
Our observations demonstrate that the cloudy structure of the 
interface region is most likely a general phenomenon, 
not restricted to the inner Galaxy.
LVCs have sizes and radial velocities
in agreement with the expectations for clouds formed in 
low-temperature fountain flows, although we measure a factor 
of ten higher HI column densities. 
Alternatively, LVCs could represent the final stages of the 
infalling intergalactic material in the on-going construction of the Galaxy. 

In the same dataset at higher negative velocities, we have
discovered a `companion' HI cloud located 50$'$ southwest of CHVC186+19-114.
CHVC186+19-114 is a typical compact high velocity cloud (HVC) 
with a well-defined core/envelope structure. 
The companion cloud has a diameter of only $7'\times9'$, and is one 
of the smallest HVCs known, most likely 
stripped from the main cloud through the interactions with the halo medium. 
\end{abstract}


\section{Introduction}

The Galactic halo, a hot, ubiquitous multi-phase gas, with a low density, and 
temperature around $10^{6}$ K, that surrounds our Galaxy is 
crucial to Galactic activity. 
As well as acting as a thermal insulator, the halo constrains 
models of the interstellar medium (ISM) in the Galactic disk 
through its impact on the  dynamics of the hot intercloud component.
From the theoretical point of view, 
the Galactic fountain model \citep{Shapiro76} was the first
to address interactions between the disk and the halo.
In this model, the hot 
gas flows from the disk into the halo, starts to cool, and falls
back onto the disk.
One of the later models, so called chimney model of the ISM \citep{Norman89},
suggested that the 
disk and the halo are connected through chimneys -- large conduits
resulting from superbubbles bursting out of the disk and forming collimated
structures that can reach heights of $\sim1$ kpc.
From the observational point of view, 
numerous ISM structures (shells, superbubbles, worms, chimneys, plumes)
were found in the vertical gas distribution of our own Galaxy 
and several nearby galaxies, and have been postulated to 
play an important role in the transfer of mass and energy
between galactic disks and their halos.
However, the details of how these structures  
replenish the hot halo gas are still not fully understood \citep{Dove00}.

In understanding the ways in which the Galactic disk and halo are related
the transition region between the Galactic thin disk and the halo,
also known as the Galactic thick disk, is of special interest. 
This region extends to heights up to 1--1.5 kpc
above the plane.
The disk/halo interface region was discovered
several decades ago \citep{Shane67,Lindblad67} and was considered to
represent a smooth envelope of neutral hydrogen (HI) surrounding the
Galactic spiral structure.
Further studies \citep{Albert83,Danly89,Lockman86}
found occasional isolated halo clouds at lower heights and
suggested that this component could be pervasive, yet patchy, and
most likely dynamically connected with the disk.

Galactic fountain and chimney models, as well as recent numerical
simulations \citep{Avillez00}, predicted the existence of 
small-scale, cloudy structure in the interface region, 
with clouds ranging in size from a few parsecs to  tens or
hundreds of parsecs. Recent observations with the Green Bank telescope
(Lockman 2002) have actually revealed dramatic small-scale structure
of the interface region in the inner Galaxy, composed of numerous
discrete HI clouds.
While this confirms the expectations from theoretical models and
simulations, the full extent and the origin of 
this cloudy component is still not understood.

Another set of attractive questions concerning the  disk/halo
interface reagion comes from the idea that the 
formation of galaxies is an ongoing process \citep{Oort66}
and that the accretion of intergalactic material (IGM) 
serves as the main source of new star 
formation fuel \citep{Maller04,Murali02}. However,  
the details of the transformation from the smooth and hot IGM on kpc scales, 
down to clumpy and cold gas on star-forming scales,
still need to be worked out.
For example, what is the internal structure of 
IGM clouds at different stages of their infall?
As searches for accretion remnants around other galaxies are difficult
due to sensitivity and area limitations 
(e.g., Fraternali et al. 2002), 
our own Galaxy is still one of the best places
to start seeking answers to these questions.

HI clouds in the Galactic halo at heights $>1$ kpc have been 
studied observationally
for several decades and are often  classified into 
several distinct groups based on how much their velocities deviate from
Galactic rotation (intermediate and high velocity clouds, HVCs), 
and their relative size (compact and extended clouds).  
Velocities reach approximate $+/- 450$ \kms~in the
LSR reference frame and their sizes vary from over 100 deg$^{2}$ to 
compact HVCs with sizes of 4--30 arcmin \citep{Putman02,Bruns04a}.  
In addition, almost all HVCs, especially the more compact ones, 
clearly show head/tail or core/envelope structures, with 
heads or cores having a narrow linewidth
while tails and envelopes having a significantly broader velocity
structure, possibly showing signs of interaction with the 
ambient medium (e.g. Br\"{u}ns \& Mebold 2004).
It is still not clear whether this
range of observational properties, especially the low-end
of the cloud size and velocity distributions, 
is caused by observational limitations.
What determines the size and morphology of the observed halo clouds?  Are
these remnants of very different physical processes, or do different
`types' of clouds represent different stages of the same phenomenon ---
accretion onto the Galaxy, for example?

The questions above are just a subset of issues that still await to be
addressed. The main purpose of this paper is to summarize the ongoing
observational efforts to understand the properties of gas clouds from the halo
to the plane by studying the disk/halo interface of our Galaxy.  
This work is inspired and enabled by the new multi-beam system, 
very recently installed at the Arecibo telescope\footnote{The Arecibo 
Observatory is part of the National Astronomy
and Ionosphere Center, operated by Cornell University under a
cooperative agreement with the National Science Foundation.},
which allows sensitive, large-scale HI surveys with 
an angular resolution of 3.5$'$.
Previous large-scale surveys of the link between the disk and the halo
have been limited by resolution (Leiden-Dwingeloo Survey, 
36$'$; \cite{Hartmann97}), or an inability to trace 
features from the halo into Galactic emission (Putman et al. 2002).
Recent high-resolution Galactic Plane 
surveys \citep{Taylor03,McClure05} have been limited to 
tracing the lower velocity and latitude
HI emission and thus also do not provide adequate information 
on the disk/halo interface.

The structure of this paper is organized as follows.
We start in Section~\ref{s:galfa_background} 
by briefly introducing the consortium for Galactic studies with the
Arecibo L-band Feed Array (GALFA) and emphasizing the importance of the 
Arecibo telescope for large-scale Galactic surveys.
Section~\ref{s:obs} describes the observing
strategy adopted for Galactic HI spectral-line surveys with ALFA.
In particular, the fast scanning mode utilizing the basket-weave technique,
and a newly developed method for obtaining reference HI spectra by
multiple frequency switching, are applicable to radio telescopes in general.
In Section~\ref{s:background} we summarize major previous 
observational studies of HI in the Galactic anti-center. 
We then present our results in Sections~\ref{s:results} and \ref{s:hvc}, 
and provide extensive discussion of observational and theoretical
issues in Sections~\ref{s:discussion-obs} and \ref{s:discussion-theory}.
We summarize our results and point out future 
studies in Section~\ref{s:summary}.

\section{GALFA --- Background and Future}
\label{s:galfa_background}

GALFA is a worldwide consortium of scientists 
(currently with 86 members) interested in 
undertaking large-scale Galactic surveys with the 
Arecibo L-band Feed Array\footnote{http://www.naic.edu/alfa/} (ALFA). 
GALFA members have been working together with the National Astronomy
and Ionosphere Center in planning and executing large 
surveys to ensure their long-reached usage and 
availability by the greater astronomical community. 
The consortium consists of three sub-consortia 
for spectroscopic HI, radio recombination line, and continuum surveys. 
It is anticipated that data products from GALFA surveys 
will be archived and easily accessible by the astronomical community.

Galactic surveys with ALFA have greatly improved angular
resolution in comparison with previous large-scale single-dish
surveys. For example, the all-sky HI survey by Bell Laboratories 
\citep{Stark92} had angular resolution of 2$^{\circ}$, 
while the HI all-sky survey with the Dwingeloo radio 
telescope (LDS; Hartmann \& Burton 1997) had 
angular resolution of 0.6$^{\circ}$. 
GALFA's angular resolution of only 3.5$'$ will be especially
appreciated at high Galactic latitudes where high resolution, 
interferometric observations are still rare.
At lower latitudes several recent, high resolution, 
surveys of the Galactic plane 
exist, or are underway \citep{Taylor03,McClure05}, with a typical angular 
resolution of 1--2$'$ and a brightness sensitivity of 1.5--3 K (per
0.8 \kms~wide velocity channels). 
To provide complete information about the
large-scale HI structure these surveys 
combine data from interferometer and single dish telescopes.

The Arecibo telescope has several unique advantages 
over all other radio telescopes:
(i) its unmatched collecting area, together with a large gain and
low-noise receivers, results in an excellent and 
unrivaled filled surface-brightness sensitivity;
(ii) this is the only radio telescope 
that can {\it simultaneously} provide both an high angular resolution, 
usually provided exclusively with interferometers, and 
a continuous range  of spatial scales  (including the largest scales, 
or so-called zero spacings), usually provided only with single dish telescopes.
This unique combination of high surface-brightness sensitivity and accurate
mapping of all angular scales above 3.5$'$ makes the
Arecibo telescope extremely valuable for Galactic science. 
In addition, ALFA, a cluster of seven receivers, has increased Arecibo's 
surveying (mapping) speed by a factor of seven, 
making the telescope a vastly more efficient 
system for covering large areas than an interferometer/single 
dish combination \citep{Dickey02}. 
In the fastest mapping mode commonly used for GALFA projects, the
mapping speed is $\sim9.4$ square degrees per hour. 
This results in a rms brightness sensitivity per beam of 
0.1 K (per 0.72 \kms~wide channels and after averaging 
both polarizations). 
For a spectral feature with a FWHM of 10--20 \kms~ our 
minimum (3-$\sigma$) detectable HI column density is
(5--11)$\times10^{18}$ cm$^{-2}$.

This paper focuses on only one of major science drivers for GALFA ---
the Galactic disk/halo interface region.
The whole science case for Galactic HI spectral-line surveys 
is rich and diverse, full description can be found in 
the GALFA White 
Paper\footnote{http://www.naic.edu/alfa/galfa/galfa\_docs.shtml}.
In addition, there are several projects planned 
for continuum and radio recombination line
surveys which will be presented elsewhere. 
Currently, several different scientific HI projects are
underway, some focusing on relatively limited areas of the sky (for
studying, for example, particular molecular clouds and their possible
connection with atomic gas), while some others requiring the coverage of
the whole sky visible from Arecibo (the study of halo cloud
properties, for example). 
The general survey philosophy has been to start
with smaller, scientifically interesting regions, apply exactly the
same observing parameters (velocity resolution, scanning mode etc.), and
eventually combine all observations to build the whole ``jigsaw puzzle'' of the
Galactic HI sky observable from Arecibo 
(declination range $-1$\degree~to 38\degree).

While large-scale Galactic HI surveys with ALFA will address numerous
outstanding questions regarding Galactic science, 
their importance for other scientific areas is also large. 
For example, X--ray observations need to be corrected for Galactic
absorption using HI column density measurements.
The small-scale structure ($<30'$) in the neutral
Galactic layer can alter significantly interpretation of X--ray 
data and needs to be accounted for \citep{Bregman03,Lockman03}.
Also, for the interpretation of UV absorption spectra, HI emission spectra
with as high angular resolution as possible, approximating a
pencil-beam toward a background source, are necessary \citep{Wakker03}. 

The large-scale HI Galactic survey with ALFA started in early 2005. 
Prior to this, in September--November 2004, 
short precursor observations were undertaken to test the instrumentation, 
observing techniques, and data reduction tools. 
One of those test runs focused on a region of
interest for studies of the Galactic disk/halo interface region, 
in the Galactic
anti-center, well known to harbor a wealth of high velocity gas. As a
demonstration of the quality of HI images obtained with ALFA we
present here observations of this region and preliminary results.

\section{GALFA Observations and Data Processing}
\label{s:obs}
\begin{figure*}
\epsscale{1.}
\plotone{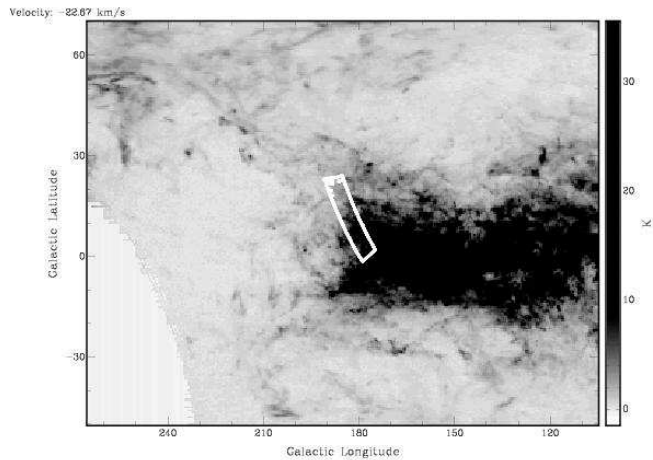}
\vspace{-0.5cm}
\caption{\label{f:ld} An image of the Galactic anti-center
region from the Leiden-Dwingello survey \citep{Hartmann97}, 
at the LSR velocity of $-22.67$ \kms, emphasizing the filamentary
structure seen at intermediate velocities. The white box outlines the
region covered in GALFA's precursor observations.}
\end{figure*}

The observations presented in this paper 
were undertaken in October and November of 2004, 
as a precursor observing run of the GALFA Consortium
with the newly installed ALFA system. 
Our main aim was testing of the newly
developed spectrometer (GALSPECT) and the proposed observing strategy.
ALFA was installed at the Arecibo telescope in April
2004.

ALFA is a cluster of seven state-of-the-art dual polarization 
receivers operating at the
frequency range 1225--1525 MHz. ALFA was constructed at the
Australia Telescope National Facility. 
Each of seven beams has a slightly elliptical shape, 
with the observed FWHM being relatively constant at 
$3.13' \times 3.57'$ (in Az and za) 
(Heiles 2004a\footnote{http://www.naic.edu/alfa/memos/}; 
Deshpande et al. in preparation).
The on-axis gain is 10.7 K Jy$^{-1}$ for the central beam and 
8.3, 8.5, 7.8, 8.3, 8.7 and 8.3 K Jy$^{-1}$ for the outer six beams, 
respectively (Heiles 2004b\footnote{http://www.naic.edu/alfa/memos/}). 
The beam efficiency (with the first
sidelobe) ranges from 0.80 for the central beam, to 0.69-0.77 for
the outer beams. The observed beam properties agree very well with the
theoretical predictions by \cite{Medellin02}. 
The central beam has the smallest and the most circularly symmetric
sidelobe; the outer beams have asymmetric sidelobes.

\subsection{GALSPECT}

GALSPECT is a dedicated spectrometer for Galactic HI
surveys. This newly developed FPGA-based
system was specifically designed for GALFA's needs. GALSPECT takes
simultaneously two spectra: a {\it science spectrum} covering 7.14 MHz with
8192 frequency channels, resulting in the fixed velocity resolution of 0.18
\kms, and a {\it calibration spectrum} covering 100 MHz with 512 frequency
channels. The usable velocity range after data processing is $-700$ to
700 \kms.

GALSPECT performs the 
frequency analysis by means of a polyphase filter bank \citep{Crochiere83}. 
This is an extension of the segmented-FFT technique 
\citep{Romney95} which improves greatly the channel 
separation (down to $-80$ dB) at a 
small increase in computational complexity (less than a factor of
two). This means that there is essentially no `spectral leakage',
which results in superb interference isolation. 
The data are digitized using 8-bit A/D converters.
With this many bits,  the clipping correction is small and we do not apply it.

\subsection{Observing strategy}

The adopted observing strategy takes advantage of 
the basket-weave (or meridian-nodding) scanning. 
We observe always at the meridian by driving the telescope only in zenith
angle. Each day a single basket-weave scan is completed.
As the Earth rotates a scan draws a zigzag pattern in the RA--Dec
coordinate frame. 
On consecutive days we obtain adjacent scans, shifted by $\sim1.79'$, 
covering eventually the whole region of interest. 
To cover the region about 30\degree$\times$5\degree~in size that is
presented in this paper, we required 2 hours $\times$ 8 days.
The basket-weave scanning technique has been commonly used in the past for
all-sky continuum surveys (for example see Haslam et al. 1982). It has
several important advantages when imaging large areas of the sky:
fast coverage of a large area of the sky, the beams 
always have the same orientation, inter-woven scans have many 
crossing points that allow a fine gain adjustment, and transit 
observations limit contamination from position-dependent sidelobes 
(which at Arecibo can be significant).
This very fast observing mode scans the sky 
at a rate $\sim5.7$ times faster than the sidereal rate, 
but covers every piece of the sky twice. The resulting integration time is
$2.45$ sec for each Nyquist pixel ($1.8'\times1.8'$ in size).
The separation between ALFA beams is $\sim1.79'$ on the sky,
close to Nyquist sampling, and it does not depend on the scanning
speed. Practical details are given in the ``GALFA HI User's
  Guide'', Kr\v{c}o et al. 
(2006)\footnote{http://www.naic.edu/alfa/galfa/galfa\_docs.shtml}.

Obtaining reference spectra is
one large challenge GALFA HI surveys have to deal with. 
While mapping large-scale Galactic structure
it is impossible to find an `empty' point on the sky that will serve
as a reference position. Commonly, the frequency switching technique 
is used when performing large-scale Galactic surveys. 
However, at many single-dish radio telescopes, and
especially at the Arecibo telescope, the classic frequency switching
technique introduces baseline problems because of frequency 
structure in the baseline produced by reflections.
We have developed a new technique for obtaining a reference spectrum, 
called ``The Least-Squares Frequency Switching'' (LSFS, Heiles 
2005a\footnote{http://www.naic.edu/alfa/galfa/galfa\_docs.shtml\#obs\_data}).
This technique derives the IF bandpass correction from a set of
observations obtained with 7 LO frequencies.
At the beginning and end of each day's basket-weave 
scan we spend about 10 minutes to obtain LSFS data. 
The LSFS technique is based on cycling among 7 different LO frequencies 
and using the least-squares technique to separate the IF gain 
from the RF spectrum. LSFS is performed separately on 
the 100-MHz calibration spectrum and on the 7.14-MHz science spectrum.  
Broad frequency structure in the IF  bandpass and
baseline of the science spectrum are derived from those portions of the
calibration spectrum that lie outside the science spectrum's band, and
appropriate corrections are applied to the science spectrum.

\begin{figure*}
\epsscale{1.}
\plotone{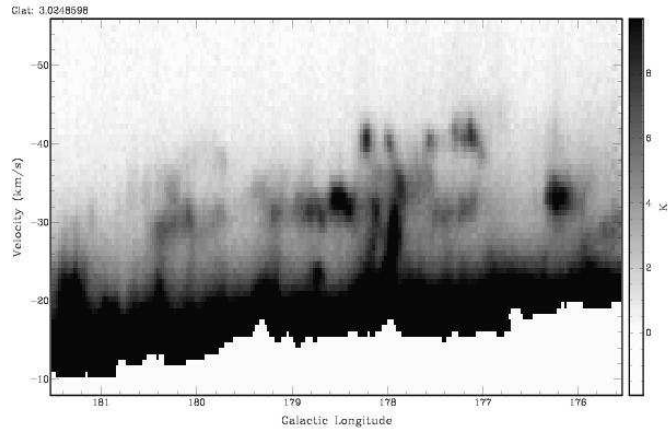}
\caption{\label{f:vel-ba} An example of interesting structure
found in the anti-center data cube: cloudy structure at $b=3$\degree. 
A chimney-like  extension is noticeable around $l\sim178$\degree~which may be
  connected with numerous small clouds at the velocity of $\sim-40$
  \kms. To enhance weak features all pixels with $T_{\rm B}>30$ K
  have been masked out.}
\end{figure*}

\begin{figure*}
\epsscale{1.}
\plotone{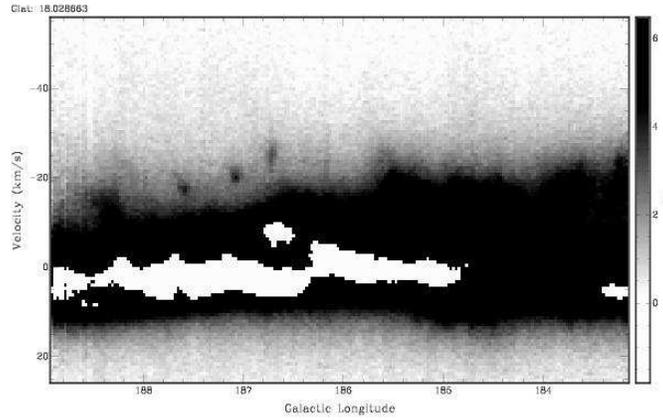}
\caption{\label{f:vel-bb} An example of interesting structure
found in the anti-center data cube: three remarkable small HI clouds 
at $b=18$\degree, $l=186$--188\degree, at the LSR velocity
  $-21$ \kms. All pixels with $T_{\rm B}>12$ K
  have been masked out. 
}
\end{figure*}

\begin{figure*}
\epsscale{1.}
\plotone{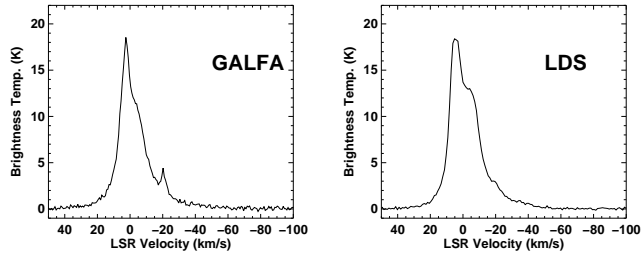}
\vspace{.5cm}
\caption{\label{f:galfa_lds}An HI velocity profile for the
  low-velocity cloud at $(l, b)=$(187\degree, 18\degree) as
  obtained with GALFA and from the LDS data. This is the middle cloud
  seen in Figure 3.
These profiles clearly demonstrate that the visibility of emission 
from a small feature at a velocity of $-20$ \kms~is dramatically 
improved by a factor of 100 smaller beam solid area achieved by GALFA.}
\end{figure*}

\subsection{Data reduction}

A special purpose suite of IDL programs has been developed for
the reduction of GALFA HI data, primarily by C. Heiles and 
J. Peek.
The data reduction consists of five stages: bandpass correction, 
temperature calibration, cross-point correlation, fixed pattern noise
reduction, and gridding.
During the first stage, spectra from all beams are corrected for
the IF gain, using the baseline as derived from the wideband calibration
spectrum. We also correct here for a fast ripple that arises
from reflections in the optical fiber that connects the control 
room with the feed cabin.

In the second stage of data reduction, spectra from all beams are calibrated into
antenna temperature units using a fiducial conversion factor, which is later  
greatly refined.
After this, crossing points of all
beams are found in the data (this is typically a large number
$\sim420/\cos({\rm Dec})$ per square degree), 
crossing spectra are compared and the
relative point-to-point gain corrections are determined.
At the fourth stage, baselines of the beams are partially corrected for
the fixed pattern noise
(Heiles 2005\footnote{http://www.naic.edu/alfa/memos/}) 
using a beam-averaging process.

Finally, the last stage of data reduction includes gridding of calibrated and
gain-corrected spectra from all scans into a spectral-line data cube.
We use a program that assumes a Gaussian convolving kernel. Prior to
gridding all data are organized in a sparse matrix form which provides
a memory-effective way of addressing a huge amount of data in IDL.
The final post-gridding angular resolution is 3.5$'$. 
In this particular data cube, a 1-$\sigma$ 
noise level is 0.16 K, per 0.74 \kms~wide velocity channel. 
The final data cube was at the end absolute-calibrated to match the brightness
temperature scale of the LDS data for this region. 
To achieve this we scaled all spectra by a constant factor of 0.84.

\subsection{Effects of stray radiation}

 The stray radiation properties of the Arecibo telescope 
are not well measured and understood. 
Above, we gave the main-beam plus first-sidelobe efficiencies; these
vary from $\sim 0.7$ to 0.8, which leaves a substantial residual beam
response from sidelobes.  A major question is: do these sidelobes reside
at large angles from the main beam --- normally referred to as ``stray
radiation'' --- as they do for smaller telescopes having significant
blockage from structures like feed legs, or are they more concentrated
toward the viewing direction?

        We have performed some rough tests to examine this question. We
compared emission profiles obtained in the Millennium survey 
\citep{Heiles03a} with the LDS data, which are corrected for most
effects of stray radiation. For data taken at modest zenith angles,
below $\sim 17^\circ$, the Arecibo and LDS profiles have similar shapes
in the line wings, where the effects of stray radiation are most
noticeable. Moreover, our comparison of Millennium survey profiles of the
same position taken at different times of year shows excellent
agreement. The absence of significant response at large angles from
beam center can be understood physically: most of the blockage at Arecibo
is produced by structures that are much larger than a wavelength, so
their diffraction patterns tend to be concentrated in a fairly tight
angle around the beam center.

        However, for pointing directions having zenith angle $\gtrsim
17^\circ$, the agreement between LDS and Arecibo profiles is not so good. Thus,
for such directions, stray radiation at Arecibo does contribute
artifacts in the line wings. This division line near zenith angle $17
^\circ$ makes sense because above that zenith angle the feed illumination pattern
falls beyond the edge of the spherical primary reflector and onto the surface
shield placed around its rim (so called skirt),
which by its very nature produces far-out sidelobes with the 
consequent stray radiation. A project for the future is to pursue 
the measurement of this stray radiation in more detail.

 Given this position-dependent stray radiation, we must ask how
significant it is for our desired science. The attractive feature of
mapping HI with the Arecibo telescope is the good angular 
resolution, compared to the LDS, combined with the high 
surface brightness sensitivity. Stray
radiation has essentially no scientific impact on this attractive
feature. The telescope's stray radiation response comes from weak sidelobes
that cover large angles in the sky, and these cannot produce structure
on  angular scales measured below a few degrees. In short, changes in
the measured HI profile on angular scales of a few degrees cannot
possibly be produced by the convolution of these weak, broad sidelobes
with structure on the sky because even for a point-source sky feature
the convolution produces a response that has the angular scale of the
sidelobes. 

        We conclude that on scales below of a few degrees, 
angular {\it structure}---i.e., {\it change} in 
the HI profile--- is not affected by
distant sidelobes.  However, on the same scales the overall shape of our
profiles are sometimes susceptible to contamination by stray radiation.
These overall shapes need to be corrected by referring to a
stray-radiation-corrected HI survey, of which the best is the
Leiden-Argentine-Bonn (LAB) survey 
\citep{Kalberla05a,Hartmann97,Bajaja05,Arnal00}. 
This survey has $\sim36'$ resolution, and when our maps 
cover a reasonable area of a few square degrees we can correct 
our profiles by using this survey as a standard.
We have not yet implemented such a correction procedure.

\section{History of studies of anomalous gas in the Anti-center region}
\label{s:background}

As mentioned in Section 2, the first GALFA observations focused on a
region in the Galactic anti-center.
The Galactic anti-center is well known as a region with a
large amount of gas at forbidden velocities and hence offers the perfect
case-study for Galactic disk/halo interfaces. 
\cite{Weaver70} was the first to point out interesting high-velocity, 
jet-like features in the Galactic anti-center region which were
apparent at some longitudes and appeared to be 
connected to the low-velocity gas in the Galaxy.
Several authors have studies these features further, 
for the full list of references please see \cite{Tamanaha97}.

The most prominent structure in the anti-center region is 
the Anti-center Shell (ACS), discovered by \cite{Heiles84}, 
and studied by several authors. This
large supershell, with an angular diameter of almost 45\degree$\times25$\degree,
is moving very fast (velocity $>70$ \kms). However it appears to have stopped
expanding. Several authors, including \cite{Kulkarni85}, have favored the
explanation that the ACS was formed by an impact of a chain of high
velocity clouds onto the Galactic disk.
\cite{Tamanaha94} pointed out that the ACS has a wealth of filamentary
structure at intermediate velocities, $-25$ to $-90$ \kms. By using a
special filtering technique he was able to isolate seven filaments and
propose a model in which filaments are located on the surface of a
sphere with a radius of about 25\degree, centered on $l=185$\degree,
$b=0$\degree. These filaments are large, having a typical length of
10\degree--40\degree, and appear at mostly constant longitudes, and have
typically FWHM of 10--20 \kms.  
The chain of HVCs in the anti-center 
region, often referred to as clouds ACI, ACII etc., has been known
for a long time. It is still unclear though whether 
these clouds are related to the ACS.

\section{The Anti-center filaments and `cold' clouds}
\label{s:results}
\begin{figure*}
\epsscale{1.2}
\plotone{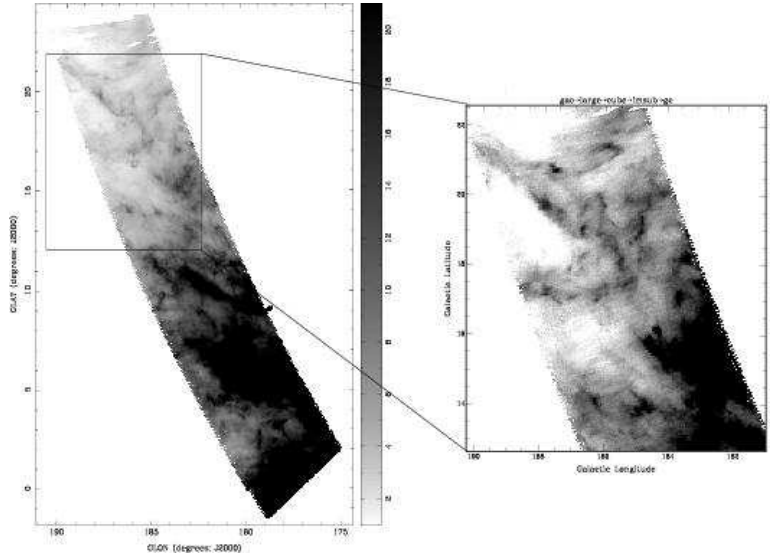}
\caption{\label{f:gac} (left) An image
obtained with ALFA showing filamentary structure at low
forbidden velocities, around $-22$ \kms. 
The grey-scale range is 1 to 21 K with a square-root transfer function
to enhance weak features. (right) A zoom in on the weak filamentary 
structure at $b\sim18$\degree~with a stretched intensity scale.}
\end{figure*}
This `pilot' HI spectral-line data cube is centered on the Galactic
anti-center region, $l\sim180$\degree, and covers a range of latitudes,
$b=$0\degree--23\degree, with $V_{\rm lsr}=90$ to $-200$ \kms.  The original
velocity resolution was 0.18 \kms, we have smoothed spectra by four
channels resulting in the final velocity resolution of 0.74 \kms.  
In the velocity domain the data cube covers several distinctly  
different environments: Galactic HI,
intermediate velocity gas at $V_{\rm lsr}\sim-50$ \kms, a portion of
the Outer Arm HVC complex at $V_{\rm lsr}\sim-80$ \kms, and a typical
compact high velocity cloud (CHVC) at $V_{\rm lsr}\sim-115$ \kms.
This CHVC was cataloged as WvW215 by \cite{Wakker91} and 
as CHVC186+19-114 by \cite{Braun99}. 

        Figure~\ref{f:ld} shows a large view of the region, obtained
        from LDS.  The white box included in this figure 
encloses the area covered in GALFA's precursor observations.  Our
data offer one of the very first views of the high-latitude HI
distribution at the high angular resolution of $\sim3.5'$ with high
surface brightness sensitivity.  This view reveals several new
and interesting phenomena, which we introduce in the following subsections.
The compact HVC is introduced in Section~\ref{s:hvc}.

\subsection{Low-velocity clouds or LVCs}

\begin{figure*}
\epsscale{1.2}
\plotone{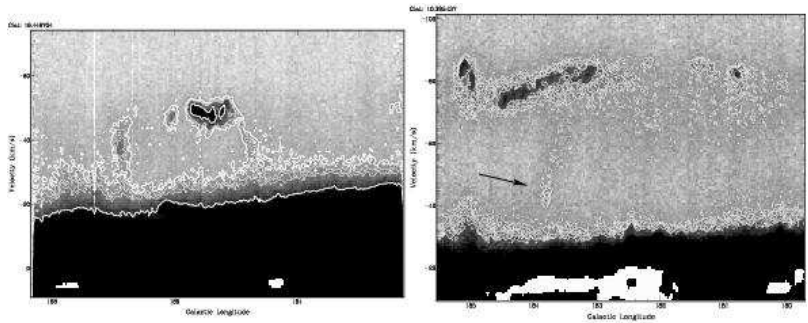}
\caption{\label{f:vel-b1} Two examples of interesting cloud morphologies
found in the anti-center data cube. 
To enhance weak features all pixels with $T_{\rm B}>30$ K
  have been masked out.
(left) An example of a larger cloud at $(l, b)=186.9, 16.4$\degree~at 
a velocity of $-38$ \kms~which shows weak elongations (tails),
  with $T_{\rm B}\sim1$ K, at both lower and higher velocities. 
Similarly, the intermediate-velocity
  cloud at a velocity of $-50$ \kms~has a tail ($T_{\rm
  B}\sim1$--1.5 K) connecting it with the disk HI emission. 
Contours are at 1.0, 1.25, 1.5 and 3.0 K. The grey-scale range is $-2$ to
  12 K with a linear transfer function.
(right) A cloud at $(l, b)=183.8, 10.4$\degree, at a velocity of
  $-43$ \kms~which appears to have a weak tail trailing all the way to
  $-80$ \kms. Contours range from 5 (0.75 K) to 10-$\sigma$ (1.5 K).}
\end{figure*}

        The region contains numerous discrete HI clouds at
negative velocities of $V_{\rm lsr} \sim-20$ to 
$-40$ \kms, shown in Figures~\ref{f:vel-ba} -- \ref{f:vel-b1}.
Throughout this paper we will call these clouds 
`low-velocity clouds' (LVCs).  
Although LVCs are in the forbidden velocity range as
the expected LSR velocity for gas in the Galactic anti-center and under
Galactic rotation is $\sim0$ \kms, velocities up to 20--30 \kms~are typically
allowed to account for turbulent motions \citep{Wakker01}.  
Most of LVCs have $|V_{\rm lsr}|$ slightly smaller than, 
or near the low velocity end of, what is traditionally assumed for intermediate 
velocity clouds (e.g. IVCs are 
defined as having $|V_{\rm lsr}|\ge30$--40 \kms; Wakker 2001, 
Richter et al. 2003). Frequently, LVCs appear to
kinematically follow the main Galactic disk structure.

        This type of HI cloud is particularly common at the lower
Galactic latitudes in our region.  An example at $b=3$\degree~is
shown in Figure~\ref{f:vel-ba}, where numerous, discrete,
narrow-linewidth HI clouds are easily noticeable.  
Some of these discrete clouds are connected in velocity to 
lower velocity gas; the strongest example is at $\ell = 178$\degree.  
In this particular data cube HI
clouds like these are not common at intermediate Galactic latitudes,
$b=\sim7 - 15$\degree, while they become again very frequent at high
latitudes, $b>15$\degree.  Clouds at high Galactic latitudes are
especially easy to notice, and most of them appear to be associated with
faint, diffuse filamentary structures. 

A spectacular example of LVCs at $b=18$\degree~is
shown in Figure~\ref{f:vel-bb}: 
three small HI clouds (the smallest cloud has an angular size of
only 4.5$'$), with extremely narrow velocity linewidths ($\sim2$ \kms) are
noticeable at distinctly different LSR velocities from those of the bulk
HI disk emission.  However, the clouds clearly follow the distribution
of the disk HI with longitude, and are probably affected by the general Galactic
rotation. Figure~\ref{f:galfa_lds} shows spectra of 
one of the three clouds as obtained with GALFA (left panel) and
from LDS (right panel). The LDS profile only barely hints at the 
$-20$ \kms~feature, which is so prominent in the GALFA profile. 
Clearly these HI clouds are too small to be visible in low-resolution 
surveys and were missed.

\subsection{Low-velocity filamentary structures}
        The region has prominent filamentary structures, especially
near $V_{\rm lsr} = -25$ \kms~--- covering roughly 
the same velocity range as that of the discrete clouds.  
These filaments are clear in Figure~\ref{f:gac} (left), which
shows a velocity channel at $V_{\rm lsr}=-22.1$ \kms.  The filaments 
may be related to the larger filaments noted previously in the 
anti-center region and studied extensively by
\cite{Tamanaha94,Tamanaha97}; however 
the character of these filaments changes dramatically  
when viewed with our full angular resolution.  
Figure~\ref{f:gac} (right) zooms in on the black square in the left
panel.  With this higher magnification, these filaments are 
peppered with numerous embedded HI clouds.  
Most of the clouds are small, $\sim5'$--10$'$ in size, 
and very difficult to notice in previous low-resolution HI images.

\subsection{Cloud morphology}

Several LVCs have especially interesting 
morphology suggestive of core/envelope structure which is 
indicative of the multi-phase medium.  
Figure~\ref{f:vel-b1} (left) shows an example of a
possible connection between the disk HI and HI clouds at 
$V_{\rm lsr}=-38$ and $-50$ \kms~--- two weak HI
`bridges' ($T_{\rm B}\sim1$ K) are noticeable extending from the disk to
the intermediate velocity gas, suggesting a possible physical connection. 
Both clouds and bridges could also represent segments of a larger, 
shell-like structure. There is no evidence, however, that the size of the
shell-like feature changes systematically with latitude, as it would
in the case of an expanding shell.
The cloud 183.81+10.37, indicated in Figure~\ref{f:vel-b1} (right) with 
the black arrow, is another interesting example --- it has a 
faint `tail' ($T_{\rm B}\sim1$ K) that trails toward  
the intermediate-velocity gas at $V_{\rm lsr}\sim-80$ \kms.

\subsection{Properties of low-velocity clouds}
\label{s:properties_lvc}
\begin{table*}
\footnotesize
\caption{Measured properties of HI clouds and filaments at low and
  high negative velocities. }
\centering
\label{table1_01}
\begin{tabular}{lccccc}
\noalign{\smallskip} \hline \hline \noalign{\smallskip}
         & Angular size    & V$_{\rm lsr}$ & FWHM   & T$_{\rm B,max}$ & $N({\rm HI})_{\rm peak}$ \\
         & (arcmin)        & (\kms)        & (\kms) & (K)             &($10^{19}$ cm$^{-2}$) \\
\hline
Low-velocity clouds & 6--12           &$-25$ to $-42$ & 3.0--7.6 &1.6--4.1      & 1--3 \\
Low-velocity filaments&a few deg &$-20$ to $-30$     &20--30    & 1--2         & 4--12  \\
\hline
CHVC186+19-114 Core & $30\times40$  & $-110$ to $-118$ &4--7 & $\sim1$--4  &$\sim1$--11     \\
CHVC186+19-114 Envelope & $30\times65$ & $-100$ to $-118$ &$\sim16$ & 0.5--1.5  &3--13   \\
CHVC186+19-114 Companion &8&$-100$ to $-107$ &$\sim15$ & 0.8  &6   \\
\noalign{\smallskip} \hline \noalign{\smallskip}
\end{tabular}
\end{table*}

To avoid confusion with Galactic emission we have selected a 
small set of 12 clouds with $b>10$\degree~which stand out 
as obvious, well-separated features. 
These clouds are distinct, well separated from the Galactic plane
emission, and have negative deviation velocities  $-42<V_{\rm dev}<-25$ \kms.  
\cite{Wakker01} defined the deviation velocity as the minimum value of
the deviation of the cloud's radial velocity from Galactic
rotation. In this framework, clouds with negative $V_{\rm lsr}$ and 
negative $V_{\rm dev}$ appear to be moving toward us too fast relative to what
is allowed by Galactic rotation.
This set of clouds is by no means complete.  
Our aim is to demonstrate the existence of small, cold HI clouds in 
the outer Galaxy, at forbidden velocities and most likely located in 
the Galactic disk/halo interface region, as well as to describe 
their basic properties.  With a larger amount of collected data 
in the near future we will be able to do a statistically more
meaningful analysis.

\begin{figure*}
\epsscale{1.2}
\plotone{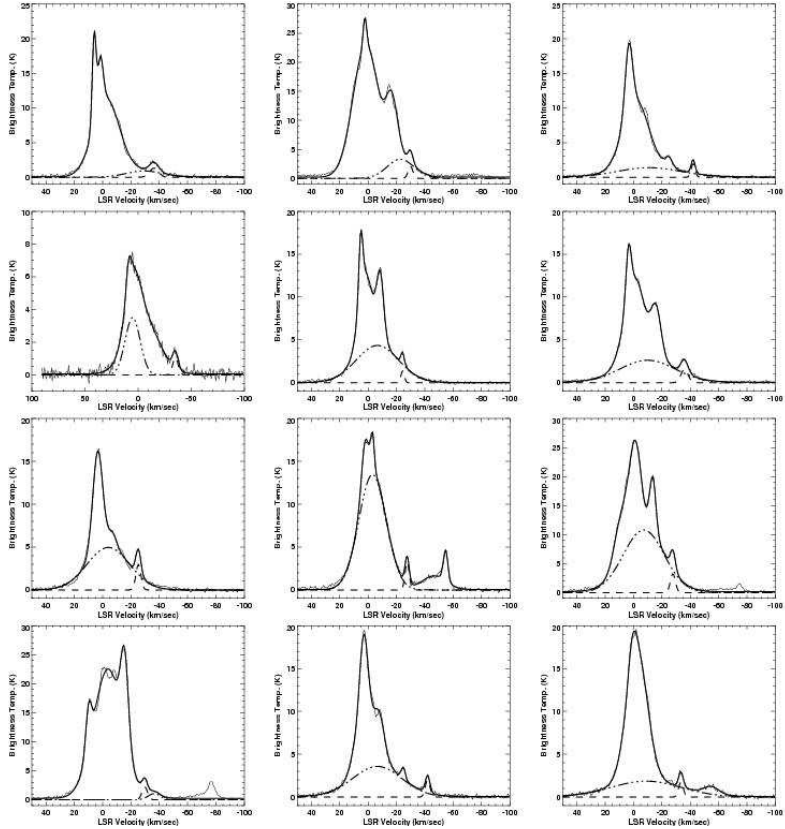}
\caption{\label{f:cores_fits} Velocity profiles of 
12 small, isolated HI clouds. The thin solid line shows the data, 
the thick solid line is the combined best fit, 
the dashed line shows the Gaussian component corresponding 
to HI low-velocity clouds, and the dot-dashed line depicts a
broad, diffuse component, which is almost always needed to produce a good fit.}
\end{figure*}

\begin{figure*}
\epsscale{1.}
\plotone{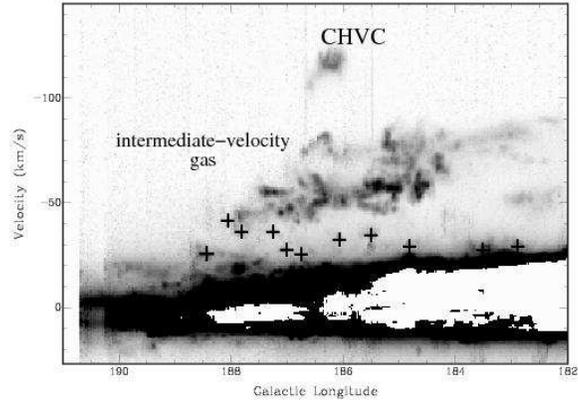}
\caption{\label{f:xpeak}An HI peak brightness temperature image
over the latitude range $b=15$\degree~to 19\degree. 
The grey-scale range is 0 to 17 K, with a linear transfer function; 
pixels with $T_{\rm B}>25$ K have been
masked out to enhance weaker features. Crosses show central
positions of selected clouds in our sample. }
\end{figure*}

Figure~\ref{f:cores_fits} shows velocity profiles through the centers of
selected HI clouds.   HI clouds/condensations easily stand out as distinct
peaks with a central velocity in the range $\sim-25$ to $\sim-42$
\kms. To measure the clouds' basic properties we have 
fitted each velocity profile
with a few Gaussian functions.  Results from this fitting procedure are
summarized in Table 1
where we provide the range of: cloud angular sizes, 
$V_{\rm lsr}$, velocity FWHM ($\Delta v$),
the peak brightness temperature (T$_{\rm B,max}$), and
the peak HI column density ($N({\rm HI})_{\rm peak}$). 
In Figure~\ref{f:xpeak} central positions of selected clouds 
are shown with crosses on a $l-V_{\rm lsr}$ image. 
Each pixel in this image represents the peak brightness 
temperature over the latitude range $b=$15\degree--19\degree. 
This image depicts almost continuous extent of cloudy structure 
between the disk HI emission and the intermediate velocity gas at $\sim-50$ \kms. 
Many of LVCs appear to kinematically follow the
disk HI, but 2--3 clouds with the most negative velocities 
could be related to the intermediate-velocity gas.
Several small clouds at positive LSR velocities, around $+20$ \kms,
are also noticeable in this figure.

        Most of LVCs must be very cold.  
The velocity FWHM for all 12 clouds is in the range 3.0--7.6 \kms.  
The mean FWHM is 4.4 \kms, and the mean $T_{\rm k,max}$ (the kinetic
temperature in the case there is no nonthermal broadening) 
is only 470 K.  80\% of clouds in this (small) sample 
have $T_{\rm k}<450$ K and FWHM of 3--4 \kms.   
The clouds' mean central LSR velocity is $-33$ \kms~and 
the mean peak HI column density is $2\times10^{19}$ cm$^{-2}$.  The
cloud angular size ranges from  6$'$ to 12$'$.  
Here we focus only on well-isolated and compact clouds;
but there are also clouds with diameters of up to 30$'$ in the data cube 
which we will discuss in a future paper.
 
\subsection{Properties of low-velocity filaments}

Most of these clouds appear related to diffuse
filamentary features (Figure 4) which are seen as broad line wings in
the Galactic HI profile. 
From the fitting of velocity profiles as well as visual inspection, 
the filaments have a large FWHM, typically
20--30 \kms, and their peak brightness temperature is low, 
$T_{\rm B}=1$--2 K. Their sizes are of order of a few degrees, some
filaments may extend beyond our maps.
Typical HI column densities are (4--12)$\times10^{19}$ cm$^{-2}$.

\section{CHVC186+19-114}
\label{s:hvc}

\begin{figure*}
\epsscale{1.4}
\plotone{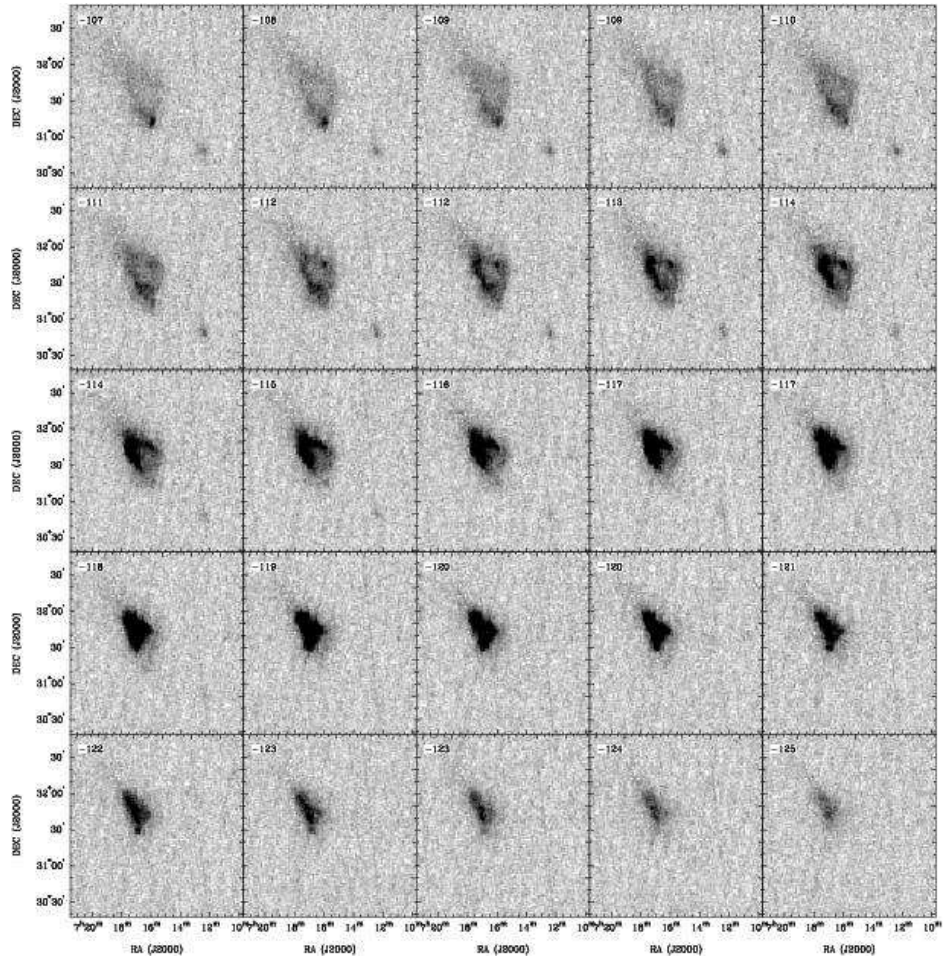}
\caption{\label{f:hvc_cube} HI images of CHVC186+19-114 at the LSR
  velocities specified in the top-left corner of each panel. The
  grey-scale intensity range is $-0.5$ to 2 K with a linear transfer
  function.}
\end{figure*}

\begin{figure*}
\epsscale{1.}
\plotone{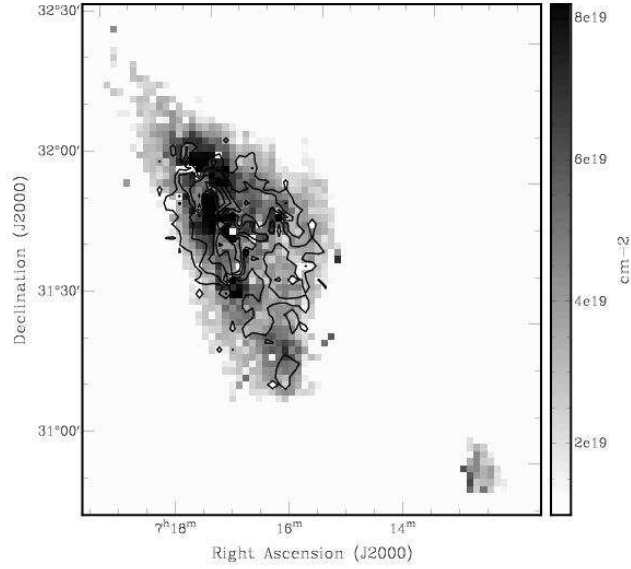}
\caption{\label{f:hvc-0mom}HI column density image of the broad
  velocity component shown in grey-scale.
HI column density distribution of the narrow velocity component is
  shown with black contours. 
Contour levels range from $1\times10^{19}$ to $1.1\times10^{21}$
  cm$^{-2}$, with a step of $2\times10^{19}$.}
\end{figure*}

\begin{figure*}
\epsscale{1.2}
\plotone{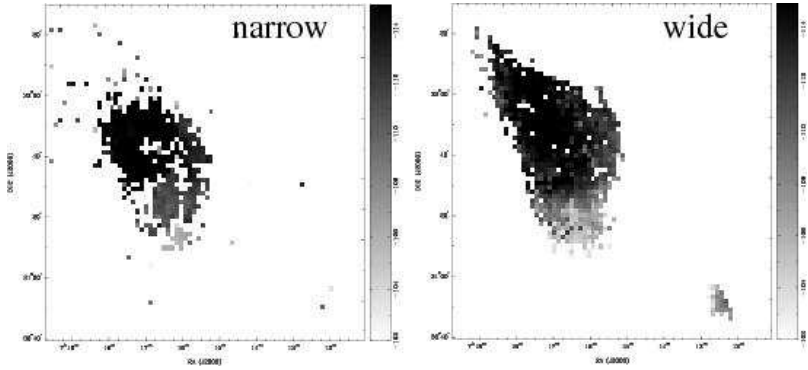}
\caption{\label{f:hvc.1+2mom}The velocity field of the narrow velocity
  component (left) and the wide velocity component (right).}
\end{figure*}

\begin{figure*}
\epsscale{1.}
\plotone{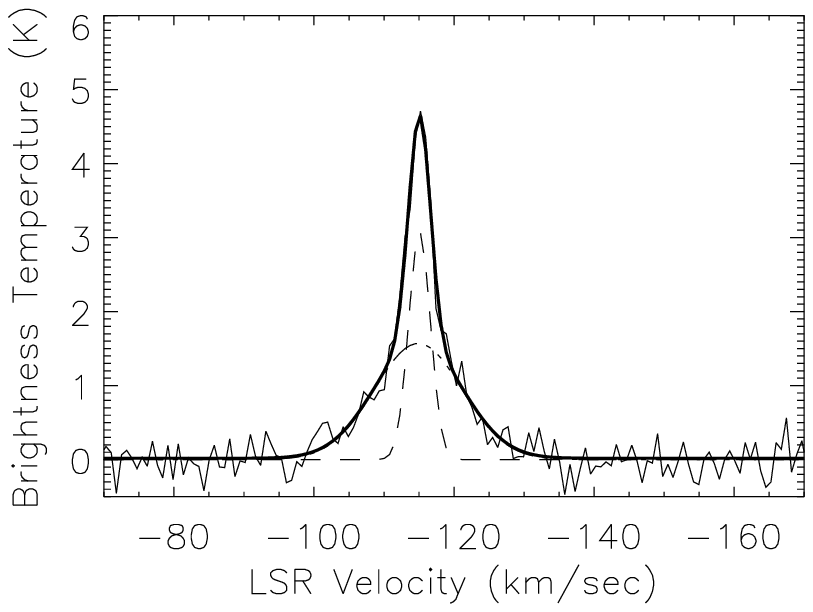}
\caption{\label{f:very_cold_clump}An example of extremely narrow
  velocity profiles found at several locations in CHVC186+19-114. 
This profile was obtained at RA 07$^{\rm h}$ 18$^{\rm m}$ 2.5$^{\rm s}$, 
Dec 31\degree 44$'$ 46.5$''$ (J2000) and was fitted with two Gaussian
functions: the narrow function has a FWHM of 3.7 \kms~(shown with the
  dashed line), and the wide function has a FWHM of 14.7 \kms~(shown
  with the dot-dashed line). The total fit is shown with the thick
  solid line.}
\end{figure*}
Figure~\ref{f:hvc_cube} shows HI channel maps of CHVC186+19-114 which
is present in the same data cube at higher negative velocities.  The top
ten panels in this figure show primarily a more diffuse, or envelope
structure, having asymmetric, ring-like morphology.  The same panels also show an
interesting `companion' HI cloud, seen for the first time in our
observations, and located $\sim50$ arcmin to the southwest of CHVC186+19-114.  
The more compact (core) component of this CHVC is primarily at
higher negative velocities, typically between $-115$ and $-122$ \kms, and
is visible in the remaining panels of Figure~\ref{f:hvc_cube}.

\subsection{Properties of core/envelope structure}

To obtain physical parameters of the HVC's core and envelope separately, we
have fitted each velocity profile with one or two Gaussian functions. For
this exercise, only pixels with a brightness temperature $T_{\rm
  B}>3\sigma$ were selected. Regions spatially corresponding to the
core structure are typically well-fitted with a combination of one narrow and
one wide Gaussian function. Most of the envelope, as well
as the HI companion cloud, is well-fitted with a single, broad
Gaussian function. Figure~\ref{f:hvc-0mom} shows HI column densities of
the narrow (contours) and wide (grey-scale) 
Gaussian components, while their velocity fields are shown 
in  Figure~\ref{f:hvc.1+2mom}. 

An interesting result is
that both narrow and wide velocity components show a clear velocity
gradient from $\sim-104$ \kms~in the south to $\sim-118$ \kms~at the
north. This suggests that the same physical mechanisms
affect and de-accelerate both cold and warm gas in this CHVC.
The central LSR velocity of the companion cloud is $-107$ \kms, 
while its typical FWHM linewidth is $\sim18$ \kms.
The wide component has a typical FWHM linewidth of $\sim15-18$ \kms,
while the narrow component has a FWHM of $\sim7$ \kms.
Interestingly, there are several small regions in this CHVC 
where the narrow velocity component has much smaller linewidth. 
As an example Figure~\ref{f:very_cold_clump} shows a velocity
profile where the colder component has a FWHM of only 3.7 \kms,
indicative of cold gas with the kinetic temperature $<300$ K.
The narrow and wide velocity components of the main cloud 
have comparable peak $N({\rm HI})$ of $\sim10^{20}$ cm$^{-2}$,
while the companion
cloud has a peak $N({\rm HI})$ of $\sim5\times10^{19}$ cm$^{-2}$.

As mentioned earlier, the main cloud of CHVC186+19-114 was 
cataloged previously \citep{Wakker91,Braun99} and also
partially imaged at high resolution with the 
Arecibo telescope \citep{Burton01} and with 
the Westerbork Synthesis Radio Telescope \citep{deHeij02}.
\cite{Burton01} found that CHVC186+19-114  has a core/envelope morphology and
fitted its velocity profiles with two Gaussian functions.
Based on the measured exponential $N({\rm HI})$ profile as a function 
of cloud's radius, being interpreted as due to a spherical exponential 
distribution of the HI volume density distribution, 
\cite{Burton01} estimated CHVC186+19-114's 
distance to be 600--800 kpc.
\cite{Maloney03}, however, showed that exponential $N({\rm HI})$
profiles could be a natural consequence of an external source of 
ionizing photons and cannot be used to derive HVC's distances.
\cite{deHeij02} noted a velocity offset between the CNM 
component, traced in interferometric data, and the
WNM component, derived from the total power, Arecibo data.
They suggested that this systematic offset 
could be explained by some type of external perturbations, 
one possibility being ram-pressure interactions with the external medium.

\subsection{Relation of main and companion clouds}

GALFA observations cover a significantly large 
area around CHVC186+19-114, revealing a `companion' HI 
cloud. The companion cloud has the central LSR velocity
very similar to that of CHVC186+19-114's envelope and 
lags behind CHVC186+19-114's core by $\sim10$ \kms.
In addition, velocity profiles ($T_{\rm B}$ and FWHM) on the
south side of the HVC envelope are similar to 
profiles through the companion cloud.
This is all suggestive of a possible physical 
association between CHVC186+19-114 and the companion cloud. 
One possibility is that this small cloud has  
been stripped from CHVC186+19-114's envelope due to ram pressure
interactions with the halo medium.
This cloud could represent an evidence for the break up of
HVCs into smaller clouds due to ram pressure (or even tidal) 
interactions. Another possibility is that both CHVC186+19-114 
and the companion cloud are embedded in lower column density 
gas and we are just able to detect the high column density peaks 
of this extended `complex'.

The companion HI cloud has an angular size of only $7'\times9'$ and 
could be classified as one of the smallest HVCs ever detected. Its size
is just at the lower end of the size distribution of the 
mini-HVC population discovered recently by \cite{Hoffman04}.

\section{Observational discussion: low-velocity clouds}
\label{s:discussion-obs}
\begin{table*}
\footnotesize
\caption{Derived (median) properties of HI clouds and filaments at low and high
  negative velocities.
}
\centering
\label{table2}
\begin{tabular}{lcccc}
\noalign{\smallskip} \hline \hline \noalign{\smallskip}
& T$_{\rm k,max}$ & $N({\rm HI})_{\rm peak}$ & $n \times D$ & $nT_{\rm k,max} \times D$\\
&  (K)    &($10^{19}$ cm$^{-2}$)     & (cm$^{-3} \times {\rm kpc}$) & ($10^{3}$ K cm$^{-3} \times {\rm kpc}$)  \\
\hline
Low-velocity Clouds & 200--1300 & 1.0--3.0 & 1.3-5.5 & 0.3-4 \\
Low-velocity Filaments& 9000--20000 & 4--12 & 0.7-1.5 & 10-20 \\
\hline
CHVC186+19-114 Core & 350--1000  &2--11 & 0.9 & 1    \\
CHVC186+19-114 Envelope  &5600  &2--13 & 0.8 & 4  \\
CHVC186+19-114 Companion&5000  &5 & 3.2 & 16  \\
\noalign{\smallskip} \hline \noalign{\smallskip}
\end{tabular}
\end{table*}

\subsection{Distances and HI masses of low-velocity clouds}

A major unknown parameter of LVCs is their distance.
As we have shown in Section~\ref{s:properties_lvc},  
80\% of LVCs in our sample have $T_{\rm k}<400$ K and 
FWHM of only 3--4 \kms. 
The peak $N({\rm HI})$ for all clouds is (1--3)$\times10^{19}$ cm$^{-2}$.
These linewidths and column densities are very similar to those 
of ordinary CNM clouds at $|b|>10$\degree~studied through 
HI absorption measurements \citep{Heiles03b},
suggesting that the anti-center LVCs
are likely to be ``true'' CNM clouds. 
As mentioned earlier, LVCs often appear
related to larger, warmer filamentary structures, with a 
typical FWHM of 20-30 \kms.

In the Galaxy, the CNM and WNM can co-exist only
over a certain, well-defined range of thermal pressures, 
$P_{\rm min}$ to $P_{\rm
  max}$ [this was first emphasized by \cite{Field69} and most recently
re-examined  by \cite{Wolfire03}].
This co-existence of cold LVCs with warm filamentary structures can be
used to estimate cloud distances.
As the ISM is highly turbulent, often turbulent pressure is more 
important than the thermal pressure. 
However, as nicely explained in \cite{Wolfire03}, on the
scales of typical CNM clouds thermal pressure dominates, and CNM
clouds are expected to be embedded in a spatially isobaric medium 
with their surface layers having
the same thermal pressure as the ambient warm medium.
\cite{Wolfire03} calculated $P_{\rm min}$ and $P_{\rm max}$ in the
Galactic midplane for several Galactocentric radii ($R_{\rm g}$). 
Both quantities decrease with
$R_{\rm g}$. For example, at $R_{\rm g}=8.5$ kpc, $P_{\rm min}=1960$ K
cm$^{-3}$ and $P_{\rm max}=4810$ K cm$^{-3}$, while at
 $R_{\rm g}=18$ kpc, $P_{\rm min}=272$ K
cm$^{-3}$ and $P_{\rm max}=1220$ K cm$^{-3}$.

We can now calculate for our LVCs their volume
density ($n$) and thermal pressure ($P$) using:
$n  \propto N({\rm HI})_{\rm peak} \times D^{-1}$ and 
$P \propto nT_{\rm k,max}$, where 
$T_{\rm k,max}=21.86 {\Delta v}^2$.
Both $n$ and $P$ depend on the cloud distance $D$, also these equations
assume that clouds are spherically symmetric 
(please note that sheets will
have higher pressure, while cigars oriented towards the observer will
have lower pressure).
Estimates for both $n \times D$ and $P \times D$ are given in Table 2,
together with the estimated maximum kinetic temperature 
and the peak HI column density.
We can now compare measured values for $P \times D$, for each cloud, with the
thermal pressure allowed for the CNM and WNM co-existence to 
estimate $D$. 
If we assume the typical thermal pressure in the Solar 
neighborhood of 3000 K cm$^{-3}$, 
then clouds in our sample have the distance range 0.1--1.3 kpc, 
with the median distance being 0.3 kpc.
The lowest thermal pressure that allows the CNM and
the WNM to co-exist is $\sim270$ K cm$^{-3}$
\citep{Wolfire03}, 
this places an upper limit on the distance to our clouds of $\sim3.6$
kpc (this value is the median for the whole cloud sample).
The highest thermal pressure that allows the CNM and
the WNM to co-exist is $\sim5000$ K
cm$^{-3}$ (for $R_{\rm g}\ga8.5$ kpc as our clouds are in the direction
of the Galactic anti-center), this places a lower limit 
on the distance to our clouds of
$\sim0.2$ kpc.
If the thermal pressure is, however, $>5000$ K
cm$^{-3}$ then cloud distances could be $<0.2$ pc and our clouds may be
very closeby. 
Such high thermal pressures are not expected to be common in the
Galaxy for $R_{\rm g}>8.5$ kpc from the theoretical point of view.
If present they would indicate the existence of very short-lived, transient clouds.
The observed morphology of LVCs, in the form of discrete, isolated
clouds with sharp edges and appearing over many velocity channels, 
does not suggest their transient character.
However, future direct distance measurements 
are essential to confirm cloud distances.
We also caution that our distances were estimated under the
assumption of the {\it local} pressure equilibrium at cloud surfaces. 
Several recent observations and numerical simulations have challenged the 
global pressure equilibrium condition for the interstellar gas, however 
still mildly favoring the local pressure equilibrium state 
(for example see Mac Low et al. 2005).

Table 2 also gives $P \times D$ for low-velocity filaments,
which appear higher than for the clouds and are most likely
due to a non-spherical, sheet-like morphology. 
This is supported by the visual appearance 
of filaments which have a far more elongated morphology than the clouds.  

If we assume for now that the LVCs have distances in the range
0.2--3 kpc, then their height from the plane
is in the range $60$--900 pc.
As the typically assumed scale height (half-thickness)
of the Galactic thin disk is $z_{\rm thin}=100$ pc
\citep{Belfort84,Kulkarni88,Avillez00},
most of the LVCs appear located in the transition
region between the thin disk and the hot Galactic halo.
With a typical angular radius of 9$'$, LVCs' 
linear size is 0.5--8 pc, while their  
HI mass ranges from $3 \times 10^{-2}$ to 7 M$_{\odot}$. 
If the clouds are embedded in the warm medium with a temperature of
6000 K, then the
mass-loss rate due to evaporation (calculated using eq. [47] in 
\cite{McKee77}) is $1.5\times10^{24}$ gr yr$^{-1}$, and hence
the cloud evaporation timescale is $\ga100$ Myr.

Several previous observational studies 
have found cold clouds at
significant heights above the disk in our own Galaxy and in several
other spiral galaxies.
\cite{Crawford02} obtained NaI and CaII observations towards the
opening of the Local Interstellar Chimney and found
cold neutral clouds with low negative velocities, located
at $0.3<|z|<2.5$ kpc. They suggested that these clouds most likely belong
to a scattered population of infalling diffuse clouds,
consistent with a Galactic Fountain model.
\cite{Howk05} discussed evidence for
cold, CNM-like, gas in the disk/halo transition region of
 several spiral galaxies based on optical imaging of extra-planar
 dust. They suggested that this gas was most likely expelled from
the thin disk through a quiescent process.
\cite{Richter03} found evidence for H$_{2}$ in dense, mostly
neutral phase, and closely linked to CNM clouds in IVCs at even higher
$z$ heights.
These results support our distance estimate, however the formation mechanism(s)
for cold clouds in the disk/halo transition 
regions remain under debate (see Section 9.1).

\subsection{Is there a link between our low-velocity clouds
and Lockman's clouds?}
\label{s:low-galactic}

{\rotate
\begin{table*}
\footnotesize
\caption{Median properties of `different' types of HI clouds in the Galactic
  disk/halo interface region. }
\centering
\label{table1_02}
\begin{tabular}{ccccccccccc}
\noalign{\smallskip} \hline \hline \noalign{\smallskip}
Reference&Number of& $l$     & $b$        & V$_{\rm dev}$ & Size & FWHM
 &$N({\rm HI})_{\rm peak}$ & $D$  & $|z|$ & $M_{\rm HI}$ \\
   & clouds      &(\degree)& (\degree)  & (\kms)        & (pc)& (\kms)
&($\times10^{19}$ cm$^{-2}$) &(kpc)& (pc)& (M$_{\odot}$)\\
\hline
Lockman (2002)&38& 28--29  &$-4$ to $-12$& 15 to 40\tablenotemark{a}     &24 &
12&2  &7.5  &950& $\la50$  \\
Stil et al. (2005)&17&18--67 &$-1.3$ to $1.3$& 25 to 60     & $\sim10$\tablenotemark{b}&$6$
&20 & 8&80& 60 \\
This work &12&183--188 & 12 to 23   &$-25$ to $-40$     & 0.5--8&4&
2  &0.2--3&60--900& $0.03-7$     \\
\noalign{\smallskip} \hline \noalign{\smallskip}
\end{tabular}

\tablenotetext{a}{One half of clouds in this study has V$_{\rm dev}\le15$
\kms. Median V$_{\rm dev}=13$ \kms.}
\tablenotetext{b}{Most of the clouds in this study appear significantly elongated.}
\end{table*}
}

One of the most important results of this study is that GALFA
observations clearly demonstrate the cloudy structure of the
Galactic disk/halo interface region in the outer Galaxy.
The cloudy disk/halo interface, emphasized recently
by \cite{Lockman02} in the inner Galaxy, is obviously
not restricted just to the inner Galaxy and could be
quite extended radially. We compare here properties
of the disk/halo interface clouds in the inner and
outer Galaxy.

Recent HI observations with the Green Bank telescope 
showed that the halo in the inner Galaxy 
is not smooth but populated with discrete HI clouds \citep{Lockman02}.
These clouds, often referred to as
``Lockman's clouds'', are located $\sim900$ pc below the plane. 
However, they appear to kinematically follow the HI distribution in the disk. 
We summarize typical properties of this cloud population in Table 3. 
Frequently, clouds are organized into larger
structures and connected by diffuse envelopes and/or filaments which
can have sizes up to $\sim1$ kpc. 
In addition, high resolution observations \citep{Lockman04} show that 
most Lockman's clouds have significant sub-structure, 
suggestive of the presence of a two-phase medium.

\cite{Stil05} discovered an additional set of 
17 isolated HI clouds in the inner Galaxy, 
with velocities up to 60 \kms~greater than the maximum 
velocity allowed by Galactic rotation (see Table 3).  
Despite their forbidden velocities,
these clouds follow the main disk emission and therefore
cannot be classified as high-velocity gas. 
Located within the Galactic thin disk, at $|z|\la80$ pc, 
these fast-moving clouds demonstrate 
that interstellar clouds can have very large 
random velocities along the line-of-sight.
Stil et al. (2005) suggested that the fast-moving clouds, at
$|z|\la80$ pc, and Lockman's clouds, at $|z|\sim950$ pc,
may belong to the same population of clouds that is widely 
spread throughout the Galaxy. Both works focused exclusively 
on clouds at tangent points in the inner Galaxy and therefore have 
well-determined cloud distances.

Table 3 lists properties of LVCs in the outer
Galaxy to facilitate their comparison with HI clouds in the inner Galaxy.
This table gives: the number of clouds in each study, the range of 
Galactic longitude and latitude,
cloud deviation velocities (as defined in Section 5.4), 
median linear sizes,
FWHM,  peak HI column densities,  typical distances,
heights above the plane, and estimated HI masses.

Several potentially important differences are noticeable.

(1) It is striking that our LVCs, which lie 
outside the Solar circle, are on average smaller, 
colder, and less massive than clouds in the inner Galaxy. 
For our selected sample, most
clouds have FWHM of $3-4$ \kms, a factor of three smaller 
than that of Lockman's clouds.
Angular sizes of LVCs are in the range 6$'$--12$'$ (with a 
few exceptional examples having a size of only 4$'$), while Lockman's
clouds have a median size of $\sim11'$. 

Linear sizes and HI masses of LVCs are of course
uncertain as they depend on cloud distances.
However distances of 6--8 kpc, which would bring HI masses
close to those of Lockman's clouds, would 
imply the thermal pressure of 100 K cm$^{-3}$ which is below 
the minimum pressure at
which the CNM can exist in thermal equilibrium \citep{Wolfire03}.
Cloud angular sizes in our study though
could be partially a selection effect, 
as we have focussed on well-isolated compact clouds for which Lockman's
angular resolution is less adequate.

(2) $|V_{\rm dev}|$ of LVCs is on average higher 
than that of Lockman's clouds.
The median deviation velocity for Lockman's sample is 
$13$ \kms, while LVCs have 
$\langle |V_{\rm dev}| \rangle=30$ \kms.
If real, this difference could be important, suggesting that
interface clouds at larger galactocentric radii ($>8$ kpc) have larger deviation
velocities than clouds at smaller radii ($\la5$ kpc).
Again, the difference in $|V_{\rm dev}|$ is at least partially due to our
selection criterion --- to avoid confusion with the disk emission 
we have selected clouds with $|V_{\rm dev}|\ga25$
\kms, but clouds with lower $|V_{\rm dev}|$ also exist in the data set.
Larger cloud samples from the GALFA surveys will be able to clarify
whether clouds with larger $V_{\rm dev}$ are more common in the outer
Galaxy.

(3)
It is also important to notice that 
LVCs have negative $V_{\rm dev}$, meaning they
are moving toward us at velocities too fast
to be explained by the Galactic rotation.
Lockman's clouds, however, have systematically
positive $V_{\rm dev}$, meaning they
are moving away from us with velocities too fast
to be explained by the Galactic rotation.

In conclusion, if our clouds would have distances 
of 6--8 kpc then their properties would be 
similar to those of Lockman's clouds, but this
would imply the thermal pressure below what is traditionally 
allowed for the CNM.
If, however, our clouds are at distances $\la2-3$
kpc, then they are colder, smaller, and less massive than
Lockman's clouds.
Taking into account various selection effects and different 
resolution provided by the Arecibo and Green Bank telescopes,
clouds in the inner and outer Galaxy may belong to 
the same population with a range of properties ($z$, sizes, HI masses).
However, distinctly different formation mechanisms
can not be excluded either at this stage.
Halo clouds with sizes similar to those of our LVCs were reported recently by
\cite{Kalberla05} and Dedes et al. (in preparation).
Using the Effelsberg radio telescope these authors
found about 20 HI clouds at $R_{\rm g}=13-17$ kpc and 
$z=3-5$ kpc. However, when viewed at
high resolution with the Very Large Array, these clouds resolve into
smaller cores with a FWHM of 3--7 \kms, a peak HI column density of a few
$\times10^{19}$ cm$^{-2}$, and diameters of a few pc.

\section{Observational discussion: Ultra-Compact HVCs}

The CHVC studied in this paper, CHVC186+19-114,
is fairly typical of the $\sim$250 currently cataloged CHVCs with
diameters generally on the order of 20$'$-80$'$ 
\citep{Putman02,Heij02a}.
The resolution and sensitivity of GALFA observations offer an unique 
opportunity to detect a large number of newly identified ultra-compact HVCs 
(UCHVCs), or HVCs with diameters $\la20'$.  
The number of these UCHVCs that have been identified to date remains limited.
The smallest was found by \cite{Bruns04a} and has an angular size of 
4.5$'$ and a peak HI column density of 
$2\times10^{20}$ cm$^{-2}$ \citep{Bruns04a}.
Slightly larger UCHVCs ($> 9'$; referred to as mini-HVCs) were found by
  \cite{Hoffman04} with much lower peak column densities ($< 10^{19}$ 
cm$^{-2}$); a similar UCHVC was mapped by \cite{Richter05}.
These UCHVCs have been completely missed in the lower
resolution surveys and it remains unclear if they will follow the power 
law distribution of flux density, column density and size 
found for other HVCs, or if they represent a distinct population of objects.

The companion cloud in Figure 8 has a size of only $7'\times9'$
and a linewidth of 18 \kms~and can thus be classified as one of the 
smallest HVCs known.  These UCHVCs may represent a variety of objects.
They may be the breaking up of a large cloud as it is stripped
via a combination of tidal and ram pressure forces.  The UCHVCs may
also represent high column density peaks in a mostly ionized or lower
column density high-velocity complex. In this case the UCHVC may 
represent the beginnings of a halo cloud 
cooling out of the hot diffuse medium. 

The close match between the
position and velocity of the UCHVC and the CHVC in our data does indicate
they are directly related.  
This is similar to the CHVCs in the vicinity
of the Magellanic Stream that show a spatial and
kinematic link to the main structure of the Stream \citep{Putman03,Bruns05}.
Also, properties of the companion cloud (angular size and FWHM) are similar to
those of HI condensations (cores) observed with the Westerbork Synthesis Radio
Telescope in the case of CHVC$120-20-443$ \citep{deHeij02}. 
CHVC$120-20-443$ is one of the only two known CHVCs 
with exceptionally broad linewidths in its CNM (core) condensations,
and \cite{deHeij02} suggest 
that it may be undergoing ram-pressure or tidal stripping.
Finally it remains possible that the 
UCHVCs represent small dark matter halos as proposed for the 
CHVCs \citep{Braun99}.
Future GALFA observations of larger areas on the sky will be
important for establishing the spatial distribution
and properties of UCHVCs.

\section{Discussion: Theoretical issues}
\label{s:discussion-theory}

\subsection{Galactic Fountain and ballistic clouds}

One of the most common models for the interaction between the Galactic
disk and the halo is based on Galactic fountain 
flows \citep{Shapiro76,Houck90}. 
In this model, hot gas from the disk is pushed up by the 
stellar activity in the form of buoyant outflows, it
travels through the halo, cools down, and as a result of thermal
instabilities rains back onto the disk in the form of cold clouds.
In their best model for low-temperature fountains (model 3; for a halo
temperature of $3 \times 10^{5}$ K and a volume density $10^{-3}$
cm$^{-3}$) \cite{Houck90} suggested that about 70 cold halo 
clouds should be present above a disk area of 400 pc$^{2}$, 
having a typical size of $\sim3$ pc, a temperature of 500 K, and a
column density of $2.5 \times 10^{18}$ cm$^{-2}$. A peak in cloud
distribution is expected at the location of cloud formation, 
$z\sim1$--2 kpc, however if drag is important then 
the density of clouds at $z<0.5$ kpc could be significant too. 
The expected radial velocities for fountain clouds are 
in the range from $-100 \sin b$ to $+50 \sin b$ \kms.
3-D simulations of a Galactic fountain by \cite{Avillez00} also show
the formation of numerous small cloudlets, with sizes down to 
their resolution limit of 1.25 pc, 
resulting from a breakup of collimated structures and also
larger clouds being swept up by blast waves.

Typical sizes of LVCs are surprisingly similar 
to what is expected for fountain clouds. 
In addition, at $b=17$\degree, the expected velocity range for 
fountain clouds is $-30$ to 15 \kms~(note that clouds at the positive
velocities would be very difficult to detect due to confusion with
Galactic HI), again very much in agreement with what we find for
LVCs. The observed $N({\rm HI})$ of
LVCs is about ten times higher that what is expected for fountain clouds, 
although \cite{Houck90} note that $N({\rm HI})$ could be higher if
clouds are concentrated at lower $z$ heights.

Another important evidence in favor of the fountain-style clouds is
hinted by cloud deviation velocities. 
\cite{Collins02} investigated a ballistic infall of fountain clouds
and showed that a vertical velocity gradient 
(with scale height $z$) is expected for
halo clouds; this is a consequence of a decrease in the 
gravitational acceleration vector with height from the disk.
Hence, at any galactocentric radius, the LSR velocity of clouds in 
the Halo, at $z>1$ kpc, will be
different from the LSR velocity of clouds in the disk.
However, the steepness of this vertical velocity gradient is a function
of the galactocentric radius at which clouds originated before 
being kicked upward from the disk.
For example, for the input circular velocity of 230 \kms~and 
at a galactocentric radius of $R_{\rm g}=8$ kpc, the expected 
$\Delta V_{\rm lsr}$ is $\la30$ \kms~when $z$
changes from $z=0$ to $z\sim2$ kpc (assuming an initial 
cloud kick velocity of 100 \kms). 
Note that viscous interactions between Halo clouds, and/or more complex
interactions between the Halo medium and clouds
can decrease $\Delta V_{\rm lsr}$.
However, for very small galactocentric radii, $\Delta V_{\rm lsr} \approx
0$ \kms, for any $z$. 
The expected LSR velocity difference ($\Delta V_{\rm lsr}$) for clouds
at higher $z$ relative to clouds in the disk, is therefore something we can
directly compare with the cloud deviation velocity, $V_{\rm dev}$.

The anti-center LVCs have  $\langle |V_{\rm dev}|
\rangle=30$ \kms, in agreement with what is expected for fountain 
clouds at heights $z=$1--2 kpc. 
The expected circulation timescale (time for clouds to return back to the
disk) for the fountain clouds is about 50 Myr \citep{Houck90,Collins02}. 
As the cloud evaporation timescale ($\ga100$ Myr) is significantly
longer than the expected circulation timescale, clouds can 
reach the disk safely before being evaporated.
In Section~\ref{s:low-galactic} we hinted at a potentially
interesting difference in $|V_{\rm dev}|$ for clouds in the inner and
outer Galaxy --- most halo clouds in the inner Galaxy appear to have smaller
$|V_{\rm dev}|$ than clouds in the outer Galaxy. 
This potential trend in $|V_{\rm dev}|$ is an important tool for
investigating cloud origin --- if halo clouds are a 
result of a Galactic fountain we would expect 
$|V_{\rm dev}|$ to increase with galactocentric radius.
Future work with larger cloud samples will investigate statistical
significance of this trend.
 
In conclusion, LVCs have typical linear sizes and
radial velocities in good agreement with expectations from the
Galactic fountain model. 
This would suggest that Galactic fountain flows, and the circulation
of fountain clouds, are (at least) as common in the outer Galaxy 
as in the inner Galaxy.
This is a surprising result when one takes into consideration 
that the Galactic star formation rate (SFR) strongly depends on galactocentric
radius, based on various observations as well as
predictions from chemical and spectrophotometric models for the
evolution of spiral galaxies (for references see \cite{Boissier99} and
\cite{Boissier03}). The star formation rate
peaks around $R_{\rm g}\sim4$ kpc and then steadily
decreases towards larger radii, at  $R_{\rm g}\sim12$ kpc for example,
SFR$\sim0.3\times{\rm SFR}_{\odot}$.
However, the decrease in SFR at large radii is alleviated partially by
the decrease of the gravitational potential which enables
vertical fountain flows to propagate easier.

\subsection{Connection between filaments and clouds}

Most of LVCs appear related to large, diffuse
HI filamentary features. Diffuse filaments and clouds both 
appear only over a restricted range of LSR velocities, 
$-20$ to $-30$ \kms.
This is suggestive of their possible physical association.

An association between larger filaments and small, 
cold clouds is expected as a result of shell or chimney fragmentation.
\cite{Norman89} expect that clouds formed from chimney walls will have
velocities $|v| \sim$10--50 \kms. 
\cite{McClure-Griffiths06} found several cold HI clouds
associated with caps of a large expanding shell and suggested that
shell fragmentation is responsible for the cloud formation.
Similarly, \cite{Koo06} proposed that small, fast-moving clouds
could be fragments of old supernova remnant shells.

Visual appearance of filaments and clouds (Figure 4) strikingly 
resembles recent numerical simulations by \cite{Audit05} of 
dynamically triggered condensation of the WNM into small CNM clouds.
In these simulations, a collision between incoming turbulent 
WNM streams creates a thermally unstable region of higher density 
and pressure but lower temperature, which starts to fragment into cold
structures. This thermally unstable WNM has a filamentary morphology,
and is seeded with numerous small (size of $\sim0.1$ pc) and cold clouds
($n\sim50$ cm$^{-3}$, $T\sim80$ K). The whole process of fragmentation
is promoted and controlled by turbulence, and
cloud morphology and properties, as well as the 
fraction of the condensed cold gas, are governed by 
turbulent properties.
Based on this apparent morphological similarity it may be reasonable
to expect that external (dynamical) forcing (through, for example,
collisions, or interactions with the dense halo gas) 
of some types of vertical WNM flows (e.g. chimneys or walls of
expanding shells) can trigger condensation of warm HI into 
cold clouds.

\subsection{Infalling multi-phase gas?}

The idea that galaxies acquire their fresh star formation 
fuel through the infall of clouds has 
attracted a lot of attention, especially recently with 
advances in numerical simulations of galaxy formation \citep{Maller04}.
As the hot gas within the cooling radius ($\sim150$ kpc for
the Milky Way size galaxy) is a subject to thermal instabilities, it
fragments into smaller, warm ($T\sim10^{4}$ K),  pressure-supported
clouds.  The warm clouds fall into the Galaxy, after undergoing
cloud-cloud collisions and ram pressure stripping, and are
tidally broken into smaller clouds before impacting the Galaxy.
The cloud formation and infall are expected to be balanced at the 
present time, and models predict abundant evidence for
this ongoing process in the Galactic halo, in the
form of numerous condensed clouds.

Kaufmann et al. (2006) investigate the formation of a galactic disk,
through N-body and SPH simulations,
and show that cloud accretion is not spherical but happens 
along the angular momentum axis. This results in outer disks preferentially
being populated with small, cold accretion remnants, 
while the central regions below and above the plane
channeling hot gas.
In this scenario, cloudy structure of the disk/halo interface 
region is expected in the inner and outer galactic disks 
but due to a very  
different phenomenon --- while outer parts are driven by
the accretion process, the inner parts could be a result of a fountain process.

However, what exactly happens to clouds during the final stages of 
their infall is still mainly unexplored, both theoretically 
and observationally.
How do large and warm clouds transform into `cloud
forms' that can easily be integrated and digested  by the disk, to
provide a smooth build up of fresh star formation fuel? 
\cite{Maller04} suggest that clouds get shredded into smaller but still
warm ($\sim10^{4}$ K) condensations,
with further cooling being stopped by the extragalactic ionizing background.
Clouds in outer disks in simulations 
by Kaufmann et al. (2006) are also warm, 100--600 pc in size, however
apparently they can cool and condense further to reach thermal
equilibrium at a temperature $<10^{4}$ K.

CHVCs and our LVCs may be able to fit into this framework 
as representing later stages of the cloud infall process.
CHVC186+19-114 shown in Figures 8--11, typical of several
hundreds of compact HVCs, could represent a cloud
that has condensed in the halo and has begun to infall, 
but has not yet reached the disk.  
The low-velocity filamentary/cloudy structure, shown in Figures 2--7, 
could represent the remaining structure of the accreted cloud 
as it is being almost integrated with the Galactic disk.
This scenario seems plausible when one considers the similarities 
between these two types of objects (see Table 1): 
CHVC186+19-114's `core' structure has velocity linewidths 
and HI column densities comparable to those found in 
LVCs, while typical HI linewidths and column densities of the 
filamentary structures at about $-25$ \kms~are very similar to 
those of CHVC186+19-114's envelope.  
If the LVCs trace the final stages of the infall process then 
they suggest cloud sizes and temperatures significantly
smaller from
what is currently found in numerical simulations.
However, details of heating/cooling and self-shielding effects 
during the cloud infall still await to be addressed.
Observationally, constraining metallicities of halo clouds would be
the crucial test of the cloud infall hypothesis.

\section{Summary and future work}
\label{s:summary}

The consortium for Galactic studies with ALFA is conducting an HI
survey of the whole Arecibo sky (declination range from $-1$\degree~to
38\degree) with high angular (3.5$'$) and velocity resolution (0.2
\kms). At the time this article is being written about 20\% of the
survey has already been completed. 
In this paper we focused only on one of the main science
drivers for GALFA --- the Galactic disk/halo interface region.
With only about 200 square degrees imaged in the Galactic anti-center
region, primarily for the purpose of hardware and software testing, 
several interesting and new phenomena have emerged.

We have found numerous small (most likely just a few pc in size) and cold 
($T_{\rm k}<400$ K) HI clouds at low negative velocities distinctly
separated from the HI disk emission (`low-velocity clouds' or LVCs). 
Over a restricted range of velocities and exactly the 
same range as for LVCs, we have also found 
larger sheet-like filamentary structures. 
Most likely, the LVCs and filaments are physically related.
While the distances to LVCs will have to be better constrained in
the future, preliminary estimates, based on the thermal pressure
requirements  for the CNM and the WNM coexistence, range from 0.2 to 3 kpc.
This implies that the clouds'  height from the plane is in the range  60 to 900
pc, suggesting that most LVCs are located in
the transition region between the Galactic disk and halo, yet they have
properties of typical CNM clouds. 
LVCs are colder and, most likely, smaller and less
massive than Lockman's clouds in the disk/halo interface region in the inner
Galaxy. In addition, LVCs may have larger absolute
deviation velocities than Lockman's clouds. 
Nevertheless, the existence of a large number of LVCs
in the outer Galaxy (at a galactocentric radius $>8$ kpc) demonstrates that
the cloudy and frothy character of the interface region is more likely
to be a general phenomenon, not restricted exclusively to the inner Galaxy.

In the same dataset but at more extreme negative velocities we
have studied CHVC186+19-114, a typical CHVC with a well-defined
core/envelope structure. As we have imaged a large area around the
CHVC, we discovered a companion HI cloud located 50$'$ southwest of
CHVC186+19-114, most likely physically 
related to CHVC186+19-114. 
Velocity profiles suggest that the companion cloud was stripped 
off the main cloud's envelope due to the interactions with the halo 
medium. This is the first time that a companion is detected 
around an isolated CHVC and may represent evidence for
the breakup of CHVCs into smaller condensations.
Based on its size, the companion cloud could be classified as a member
of the population of UCHVCs, discovered only very recently.

We have discussed briefly three possible scenarios for the formation and 
maintenance of the clumpy Galactic disk/halo interface region.
The first scenario is provided by Galactic fountains, buoyant hot flows
from the disk which cool and rain back in the form of cold clouds. 
Observational properties (cloud size and radial velocity) 
of LVCs are in a good agreement with predictions of 
low-temperature fountains models.
The second scenario is motivated by numerical simulations by \cite{Audit05} of
dynamically triggered condensation of the WNM into small CNM
clouds. Vertical WNM flows propagating upward from the Galactic plane,
in the form of expanding shells or chimneys, could be triggered into
formation of CNM clouds at larger $z$ heights. This comparison is
primarily based on morphological resemblance at this stage as detailed
models do not exist. 
Finally, the third scenario investigates whether
the low-velocity clouds/filaments could represent the final stages of 
the infalling IGM. This has been encouraged by the recent models 
and simulations \citep{Maller04,Kaufmann06} which emphasize the
multi-phase and multi-scale character of the galaxy formation
process. While simulations show encouraging qualitative resemblance at
this stage the detailed physical processes during the final stages of
cloud infall are largely unexplored. We hope that our current, and especially
future high-resolution datasets, will stimulate and guide further
theoretical and numerical work in this area.

Future GALFA datasets will clearly find more disk/halo interface clouds
and new HVCs, especially in the regime of the smallest HVCs 
which is largely unexplored.
With the larger data sets we will be able to confirm whether 
the cloudy structure exists at other Galactic longitudes, 
and also the statistical significance of the potential 
differences between interface clouds in the inner and outer Galaxy.
Also, future GALFA datasets will be 
important for establishing spatial distribution
and properties of the smallest HVCs.

\begin{acknowledgements}
It is a great pleasure to thank all members of the Arecibo Observatory
for the successful instalation and commissioning of ALFA, as well as
their support in undertaking large-scale surveys with this
instrument. We would especially like to thank Jeff Hagen and 
Mikael Lerner for the implementation of the basket-weave scanning mode as a 
standard data-taking procedure, Arun Venkataraman for 
numerous network and disk-space related endeavors, 
Eddie Castro for working very hard on many aspects of the front-end 
system, and Phil Perillat for crucial details of software development. 
We would also like to thank Henry Chen for building the GALFA spectrometer.
We are indebted to telescope operators for their superb help in 
running many GALFA projects.
The authors are thankful to an anonymous referee 
for insightful comments that improved the paper.
The GALFA Consortium web page is
http://www.naic.edu/alfa/galfa/ .
SS and CH acknowledge partial support from NAIC as Visiting Scientists
during the period of this work. 
We acknowledge support by NSF grants AST 04-06987, 00-97417, and 99-81308.
JJL was supported by the Korea Science and Engineering 
Foundation Grant ABRL 3345-20031017.
The data cube used in this study is available at
http://astro.berkeley.edu/\~{}sstanimi/Data/ .
\end{acknowledgements}

\label{lastpage}
\end{document}